berman2.tex
----------
X-Sun-Data-Type: default
X-Sun-Data-Name: berman2.tex
X-Sun-Charset: us-ascii
X-Sun-Content-Lines: 221

\documentstyle[12pt]{article}
\newcommand{\be}{\begin{equation}}
\newcommand{\ee}{\end{equation}}
\begin{document}

\begin{center}
{\large\bf Onset of Chaos in a Model of Quantum Computation}\\ \ \\
{G.P.Berman$^{[a]}$,F.Borgonovi$^{[b,c]}$,
F.M.Izrailev$^{[d]}$, V.I.Tsifrinovich$^{[e]}$ }
\end{center}
$^{[a]}$Theoretical Division and CNLS, Los Alamos National
Laboratory, Los Alamos, NM 87545, USA\\ $^{[b]}$Dipartimento di
Matematica e Fisica, Universit\`a Cattolica, via Musei 41, 25121
Brescia, Italy  \\ $^{[c]}$ I.N.F.M., Gruppo Collegato di Brescia
and I.N.F.N., Sezione di Pavia   Italy\\ $^{[d]}$Instituto de
Fisica, Universidad Autonoma de Puebla, Apdo. Postal J-48, Puebla
72570, Mexico\\ 
$^{[e]}$ IDS Department,
Polytechnic University, Six Metrotech Center, Brooklyn, NY 11201\\ \ \\

Recently, the question of a relevance of {\it
quantum chaos} has been discussed in applications to quantum
computation \cite{dima,vict}. Indeed, according to the general
approach to closed systems of finite number of interacting
Fermi-particles (see, e.g. \cite{alt,I01}), as the interaction between {\it quibits} increases a kind of chaos is expected to
emerge in the energy spectra and structure of many-body states.
Specifically, the fluctuations of energy levels and components of
the eigenstates turn out to be very large and they are described by the
random matrix theory. Clearly, if this happens in a quantum
computer, it may lead to a destruction of the coherence (due to an internal decoherence inside many-body
states) required for quantum
computations.
It is important to stress that the quantum chaos occurs
not only in the systems with {\it random} interactions, but also
for purely {\it dynamical} interactions. In the latter case, the
mechanism of chaos is the non-linear 
two-body interaction represented in the basis of non-interacting
particles.

Numerical analysis \cite{dima} of the simplest model of a quantum
computer (2D model of $1/2$-spins with a random interqubit
interaction $J$) shows that as the number, $L$, of
qubits increases, the chaos threshold $J_{cr}$ {\it decreases} as
$J_{cr}\propto 1/L$. Consequently, it was claimed that the onset
of quantum chaos is a real danger for the quantum computers with
large $L\gg 1$. On the other hand, in \cite{vict} is was argued
that in order to treat this problem properly, one needs to
distinguish between the chaotic properties of {\it stationary states} and
 the dynamical process of quantum computation.

Below, we report our theoretical and numerical results for
a realistic model of quantum computer, described in \cite{gena,qcomp}.
We consider {\it both} stationary and dynamical approaches to the
model in the region of a {\it non-selective} excitation which
prepares a homogeneous superposition of $N=2^L$ states needed to
implement both the Shor and the Grover algorithms.

The model describes a 1-dimensional chain of $L$ interacting
$1/2$-spins in the constant magnetic field $B^z$, subjected to a
sum of $p=1,...,P$ time-dependent rectangular pulses of a circular
polarized magnetic field rotating in the $x,y$-plane \cite{BDMT98,BDLT00}. Each of the
pulses has the amplitude $b^p_\perp$, frequency $\nu_p$, phase
$\varphi_p$, and is non-zero during the time $T_p=t_{p+1}-t_p$.
The Hamiltonian has the form,
$$
{\cal H}= -\sum\limits^{L-1}_{k=0} (\omega_kI^z_k+2 \sum\limits_{n
> k}J_{k,n} I^z_k I^z_n)-
$$
\begin{equation}
{\frac{{1}}{{2}}}\sum\limits_{p=1}^{P}\Theta_p(t)\Omega_p
\sum\limits_{k=0}^{L-1} \Bigg(e^{-i\nu_p t-i\varphi_p}I^-_k+ e^{i\nu_p
t+i\varphi_p}I^+_k\Bigg),
\label{ham00}
\end{equation}
where the ``pulse function" $\Theta_p(t)$ is $1$ during the
$p$-th pulse. The quantities $J_{k,n}$ are the constants of Ising
interactions between two qubits, $\omega_k$ are the frequencies of
the spin precession in the $B^z-$field, and $\Omega_p$ is the Rabi
frequency corresponding to the $p$-th pulse. The operators
$I_k^{\pm}=I^x_k \pm iI^y_k$ are defined by the relations
$I_k^{x,y,z} = (1/2) \sigma_k^{x,y,z}$, the latter being the Pauli
matrices.

The Hamiltonian for a single pulse can be written in the
coordinate system, rotating around $z$-axes with the frequency
$\nu_p$. Thus, for one pulse the model is described by the {\it
stationary} Hamiltonian (below, $\varphi _p=\pi /2,
\Omega _p=\Omega, \nu _p=\nu $). We mainly study the
nearest-neighbor interaction ({\it N-interaction}) between qubits
for the {\it dynamical} case, $J_{k,n}=J\ \delta _{n,k+1}$, and
when all $J_{k,k+1}$ are random. In contrast to the model
with homogeneous magnetic field \cite{dima}, we consider a
constant gradient magnetic field with linear dependence on the
position of the $k$-th qubit, $\delta_k=|\omega_{k+1}-\omega_k|\ll
\omega _k=ak$, with $\Omega_p\ll
J_{k,n}\ll\delta\omega_k\ll\omega_k$. Thus, for the dynamical
$N-$interaction the Hamiltonian reads,
\begin{equation}
\label{ham0}H=\sum_{k=0}^{L-1} \Big [-\delta_kI^z_k+ \Omega I^y_k\Big]
-2J\sum_{k=0}^{L-2} I^z_k I^z_{k+1};\,\,\,\,\,\,\,\,\,\,\,\,
\delta_k=\omega_k-\nu.
\end{equation}

For this Hamiltonian we have developed a theory
\cite{gena,qcomp} which predicts two transitions which depend on
the interaction $J$. The first transition was called in \cite{qcomp} the
{\it delocalization border} which corresponds to the transition to
{\it weak chaos} for,
\begin{equation}
J>J_{cr}\approx \frac{4a^2}\Omega.
\label{Jcr}
\end{equation}
By weak chaos we mean a kind of randomness in many-body
states, together with the absence of the Wigner-Dyson (WD)
distribution $P(s)$ for the spacings between energy levels of the
Hamiltonian (\ref{ham0}). The latter distribution is a strong
evidence of quantum chaos in the energy spectra of chaotic
quantum systems. It typically emerges above the delocalization
border \cite{I01}. Instead, the form of $P(s)$ in our model is
very close to Poisson, which is known to occur in integrable
systems. Our analytic approach allows one to explain this
unexpected result by showing that, indeed, the model (\ref {ham0})
is close to the integrable one, even in the case of a completely
random $N$-interaction \cite{qcomp}.

The estimate (\ref{Jcr}) turns out to be very different from that
obtained in \cite{dima} for a homogeneous magnetic field.
Indeed, according to (\ref{Jcr}), the (weak) chaos border is
independent of the number of qubits. Therefore, a magnetic field
with a constant gradient strongly reduces the unwanted effects of 
quantum chaos. Numerical data show that one needs to have a
relatively weak interaction, $J\ll J_{cr}$, in order to avoid large
errors in the structure of many-body states, which appear as a
result of weak chaos.

Another unexpected analytical prediction which is confirmed by the
numerical data, is that the delocalization border (\ref{Jcr})
remains the same for the case when {\it all} qubits interact
with each other with random interactions, $J_{k,n}$. However, in this
case, the delocalization border (\ref{Jcr}) coincides with the
onset of {\it strong chaos}. The latter is characterized by strong
(almost Gaussian) fluctuations of the components of eigenstates,
and by a WD-distribution for $P(s)$. Theoretical analysis shows
that the mechanism of this transition to strong chaos is related
to a strong overlap of energy bands in the spectra of the
Hamiltonian (\ref{ham0}).

We have also studied the errors that arise when preparing the
uniform many-body state from the ground state. For this, we computed
the evolution of the wave function in the model (\ref{ham00}),
during one pulse with $\varphi=\pi/2$. Without the interaction,
$J=0$, and at the absence of the magnetic field gradient, at the end of the pulse all components of the wave function
are the same, $\psi_n^0=1/\sqrt{N}$. The interaction causes in
some errors which can be characterized by the amplitude,
$\eta=\langle||\psi_n|-\psi_n^0|\rangle_n$, and the phase, $\phi = \langle \arctan ( Im \psi_n / Re \psi_n ) \rangle_n$,
where $\langle ... \rangle_n$ means the average over different $n$ components.
Numerical data show that the errors
decrease with an increase of $\Omega$ as $\eta \propto
\Omega^{-2}$, and $\phi \propto \Omega^{-1}$
 in agreement with simple analytical estimates. As
one can see, the delocalization border does not influence the
errors. This means that weak chaos is not important for this
kind of evolution (``non-selective'') of our system. Indeed, this evolution lasts
only a short time ($\tau=\pi/\Omega$) compared with the inverse distance between
nearest levels inside the energy band. Therefore, when the bands
are non-overlapped, weak chaos does not influence the
dynamics. On the other hand, as $\Omega$ decreases, the
bands start to overlap which strongly increases errors in the wave
function.

The work of GPB and VIT was supported by the Department of Energy
(DOE) under the contract W-7405-ENG-36, by the National Security
Agency (NSA) and by the Advanced Research and Development Activity
(ARDA). FMI acknowledges the support by CONACyT (Mexico) Grant No.
34668-E.


\begin{thebibliography}{99}
%
\bibitem{dima}  B. Georgeot and D.L. Shepelyansky, Phys. Rev. E.,
{\bf 62}, 3504 (2000); ibid, 6366.

\bibitem{vict}  V.V. Flambaum. Aust. J.Phys. {\bf 53}, N4, (2000);
P.G. Silvestrov, H. Schomeraus, and C.W.J. Beenakker,
quant-ph/0012119.

\bibitem{alt}  B.L. Altshuler, Y. Gefen, A. Kamenev and L.S. Levitov,
Phys. Rev. Lett., {\bf 78}, 2803 (1997); V.V. Flambaum and F.M.
Izrailev, Phys. Rev. {\bf E 56}, 5144 (1997).

\bibitem{I01} F.M. Izrailev, to appear in Proceedings
of the Nobel Simposia ``Quantum Chaos Y2K'', Physica Scripta, 2001;
cond-mat/0009207.

\bibitem{gena} G.P. Berman, F. Borgonovi, F.M. Izrailev, and
V.I. Tsifrinovich, quant-ph/0006095.

\bibitem{qcomp} G.P. Berman, F. Borgonovi, F.M. Izrailev, and
V.I. Tsifrinovich, quant-ph/0012106.

\bibitem{BDMT98} G.P. Berman, G.D. Doolen, R. Mainieri, and
V.I. Tsifrinovich, {\it Introduction to Quantum Computers}, World
Scientific Publishing Company, 1998.

\bibitem{BDLT00}
G.P. Berman, G.D. Doolen, G.V. Lopez, V.I. Tsifrinovich, Phys.
Rev. B, {\bf 61}, 2305 (2000); Phys. Rev. A, {\bf 61}, 2307
(2000).
%
\end{thebibliography}
\end{document}